# Physics–Informed Neural Networks (PINNs) as intelligent computing technique for solving partial differential equations: Limitation and Future prospects

Weiwei Zhang[1, 2, 3, *], Wei Suo[1, 2, 3], Jiahao Song[1, 2, 3], Wenbo Cao[1, 2, 3]

1. School of Aeronautic, Northwestern Polytechnical University, Xi' an, 710072, China;

2. International Joint Institute of Artificial Intelligence on Fluid Mechanics, Northwestern Polytechnical University, Xi' an, 710072, China;

3. National Key Laboratory of Aircraft Configuration Design, Xi' an, 710072, China

* Corresponding author. Email: aeroelastic@nwpu.edu.cn

**Abstract**

In recent years, Physics-Informed Neural Networks (PINNs) have become a representative method for solving partial differential equations (PDEs) with neural networks. PINNs provide a novel approach to solving PDEs through optimization algorithms, offering a unified framework for solving both forward and inverse problems. However, some limitations in terms of solution accuracy and generality have also been revealed. This paper systematically summarizes the limitations of PINNs and identifies three root causes for their failure in solving PDEs: (1) Poor multiscale approximation ability and ill-conditioning caused by PDE losses; (2) Insufficient exploration of convergence and error analysis, resulting in weak mathematical rigor; (3) Inadequate integration of physical information, causing mismatch between residuals and iteration errors. By focusing on addressing these limitations in PINNs, we outline the future directions and prospects for the intelligent computing of PDEs: (1) Analysis of ill-conditioning in PINNs and mitigation strategies; (2) Improvements to PINNs by enforcing temporal causality; (3) Empowering PINNs with classical numerical methods.

**Introduction**

Partial differential equations (PDEs) play an increasingly important role in modern science, revealing the mathematical laws underlying physical systems. They are crucial in analysis, prediction, and control across many fields[1]. Due to the difficulty in analytically solving most PDEs, researchers have proposed various numerical methods for solving PDEs in the past few decades, including finite difference method[2], finite volume method[3], finite element method[4], and spectral method[5]. These methods have been widely applied in many physical and engineering fields. However, these numerical methods also have some limitations, including relying on meshes[6, 7], facing the curse of dimensionality when solving high-dimensional problems[8, 9], design difficulties and weak robustness of high-order schemes[10, 11].

In recent years, the rapid development of deep learning has made it possible to solve PDEs by artificial intelligence. In 2019, Raissi et al.[12] proposed physics-informed neural networks (PINNs) for solving forward and inverse problems involving partial differential equations. The central idea of this method is to encode PDEs into the loss function of the neural network using automatic differentiation, minimizing the PDEs residuals while the network approximates the initial and boundary conditions (IBCs) and observed data. In fact, the concept of PINNs can be traced back to some researches[13-15] on solving PDEs based on neural networks in the 1990s. Since the introduction of PINNs, researchers have made significant efforts in algorithm improvements and practical applications. The growth of the number of PINNs-related papers with the topic "Physics-informed neural networks" in the Web of Science database is shown in Figure 1. PINNs have been demonstrated for various phenomena, including fluid mechanics[16-20], heat transfer[21-24], fluid-structure interaction[25-27], electromagnetic propagation[28, 29] and quantum computing[30, 31].

PINNs approximate solutions to PDEs by optimizing neural network parameters, which is notably different from classical numerical methods that solve PDEs through numerical discretization and iterative method. The fundamental idea behind classical numerical methods is to approximate PDE solutions over continuous space by piecewise continuous polynomials in discretized space. This typically involves domain discretization, equation discretization, and solving systems of algebraic equations. In contrast, vanilla PINNs use differentiable continuous neural networks to approximate the solution over continuous space. The general procedure for solving PDEs with PINNs includes neural network architecture design, weighted loss function design, and parameter optimization of the neural networks. A comparison of PINNs and classical numerical methods is presented in Table 1[32-34], which highlights the main advantage of PINNs: they leverage the flexibility and adaptability of neural networks in handling optimization problems, making it easier to integrate data (e.g., experimental observations or optimization objectives) and IBCs. This facilitates a unified framework for solving forward and inverse problems. Additionally, the well-developed code ecosystem for neural networks simplifies the implementation and maintenance of PINNs. By using scattered points to discretize the solution domain, PINNs also reduce the high discretization cost associated with complex geometries. A promising application based on the advantages of PINNs is flow visualization technology[16, 35, 36], also known as data assimilation[37-40], where the flow field or other unknown parameters can be inferred from sparse observations (e.g., concentration fields and images). This type of inverse problem is challenging for classical PDE numerical solvers. Recent studies[41-44] have also demonstrated the advantages of PINNs in inverse design and optimal control, where solving PDEs and obtaining optimal designs are carried out simultaneously. In contrast, the direct adjoint looping (DAL) approach requires hundreds of repeated solutions to the forward problem, resulting in prohibitively high computation costs for PDE-constrained optimization problems. PINNs-like methods have also shown significant results in addressing parametric problems[45-47].

Although the neural network-based optimization approach for solving PDEs gives PINNs high flexibility, it also brings certain challenges and limitations. As shown in Table 1, compared to the well-established iterative approach in classical numerical methods, PINNs solve PDEs under given IBCs through optimization. Due to the incomplete development of optimization theory, PINNs can only guarantee locally optimal solutions, which limits their accuracy. Studies indicate that the loss functions in PINNs are high-dimensional non-convex functions, making it difficult to solve with existing optimization methods. As a result, PINNs are less accurate and efficient than high-order Computational Fluid Dynamics (CFD) schemes in forward problems, and cannot replace current CFD methods[33]. Additionally, the local optimum issue also means that vanilla PINNs cannot strictly enforce IBCs. During the weighted design of the loss function for IBCs and PDEs, PINNs are highly empirical and lacks strong generality.

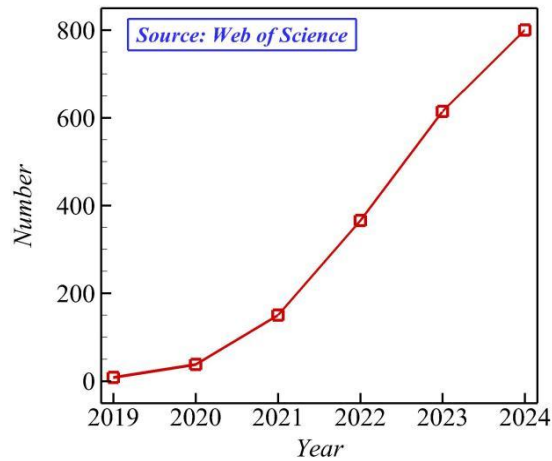

Figure 1. Number of PINNs related papers published from 2019 to 2024.

Table 1. Comparison of PINNs and classical numerical methods.

|  | PINNs | Classical numerical methods |
|---|---|---|
| Combination of data and algorithms | Easy to implement, requiring only additional constraint terms in the loss function | Difficult to implement |
| Inverse problem-computation cost | Unified framework for forward and inverse problems, no additional code required | Framework requires additional extensions, resulting in high computation cost |
| Difficulty of discretizing complex geometries | Only requires generating scattered points, no explicit mesh generation needed | Complex geometries have high mesh requirements, relying on mesh generation techniques |
| Code implementation and maintenance | Easy to implement | Difficult to implement |
| Forward problem-solving approach | Optimization-based, theory still developing | Iteration-based, well-established theory |
| Forward problem-computation accuracy | Can only guarantee local optimum, with accuracy limitations | Accuracy progressively improves with increased grid numbers or polynomial order within elements, supported by well-established approximation theory |
| Generality and robustness | IBCs as soft constraints, highly empirical, weak generality | Robust method with strong generality |

Accurately identifying the limitations of PINNs and proposing feasible improvements are crucial issues for the advancement of PINNs. Based on a review of relevant literature summarizing the shortcomings of PINNs, this paper further explores the root causes of these limitations through detailed mathematical analysis and case studies. Building on this, several recommendations are proposed to improve PINNs. Finally, the paper discusses future directions and visions for intelligent methods in solving PDEs.

**1 Discussion on the limitations of PINNs**

Currently, with the growing popularity and in-depth research on PINNs, the existing shortcomings of PINNs have been summarized in related studies, such as difficulties in loss function weight design and low solving efficiency. Based on these reviews, this paper summarizes three main limitations of PINNs: (1) Low solution accuracy and high computation cost; (2) Poor generality; (3) Numerous empirical parameters and strong parameter sensitivity. The following sections provide a detailed discussion of these three limitations.

**(1) Low solution accuracy and high computation cost**

Jagtap et al.[48] identified two main limitations in solving PDEs with PINNs. The first is solution accuracy: due to the local optimum resulting from solving high-dimensional non-convex optimization problems, the absolute error of PINNs solutions is difficult to reduce below the $10^{-5}$ level. The second is computation cost: optimizing the parameters of deep neural networks with calculating the loss of PDEs through automatic differentiation incurs high training costs. McGreivy and Hakim[49] mentioned that researchers conducting studies on PINNs must be aware of these two limitations.

Grossmann et al.[50] compared the computation time and accuracy of the finite element method (FEM) and vanilla PINNs in different types of PDEs, including elliptic, parabolic, and hyperbolic equations. The computation time refers to the time it takes for FEM and PINNs to solve the PDEs on given grids and points, while the accuracy is measured by the $L_2$ relative error. For example, in the case of elliptic equations, Figure 2 shows the curve of computation time (in seconds) versus error, using one-dimensional, two-dimensional, and three-dimensional Poisson equations for comparison. It can be observed that in terms of computation time, FEM has an advantage of 1-3 orders of magnitude over PINNs, with this advantage being more significant in lower-dimensional PDEs. As the dimensionality of the PDEs increases, the computation time advantage diminishes somewhat because the number of discrete elements required for FEM grows exponentially with the dimension of the PDEs, facing the curse of dimensionality. In contrast, PINNs can represent functions in higher-dimensional spaces without a significant increase in the number of parameters. However, overall, due to the extensive parameter updates required during neural network training, the computation time of PINNs is still much higher than that of FEM. In terms of computation accuracy, for the given maximum grid size, the highest accuracy of FEM in the one-dimensional Poisson equation is two orders of magnitude higher than that of PINNs. Moreover, similar to computation time, as the dimensionality of the PDEs increases, the difference in maximum accuracy between FEM and PINNs decreases. It is also noticeable that for smaller grid sizes, the computation accuracy of PINNs can outperform that of FEM. Additionally, for numerical methods, an increase in grid number leads to improved accuracy, while in PINNs, accuracy improvement is not directly related to the increase in the number of network parameters.

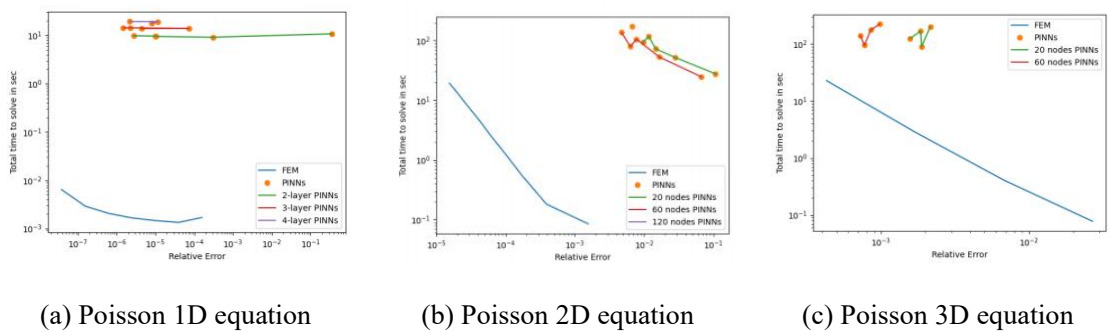

(a) Poisson 1D equation    (b) Poisson 2D equation    (c) Poisson 3D equation

Figure 2. The curve of computation time (in seconds) versus error for elliptic equations[50].

Dong et al.[51] compared the computation accuracy and time of the deep galerkin method (DGM), PINNs, and FEM, where DGM is another neural network-based method for solving PDEs. Table 2 presents the comparison of the maximum error, root mean square error, and computation time between PINNs and FEM for the one-dimensional nonlinear Helmholtz equation. In this case, PINNs used Adam and L-BFGS as optimizers, while FEM used different numbers of discrete elements. It can be seen that the maximum error of PINNs is six orders of magnitude higher than that of FEM, and in terms of computation time, PINNs require over 31.9 times more time than FEM (578.2 / 18.1 ≈ 31.9), which is attributed to the tens of thousands of epochs required for training PINNs. Table 3 compares the maximum error, root mean square error, and computation time between DGM and FEM for the one-dimensional viscous Burgers' equation. Similar to PINNs, FEM shows a significant advantage over DGM in both computation accuracy and cost.

Table 2. Comparison of error and time for solving one-dimensional nonlinear Helmholtz equation[51].

| method | elements | maximum error | rms error | epochs | training time/ wall-time(seconds) |
|---|---|---|---|---|---|
| PINN (Adam) | - | 4.56e - 3 | 5.04e - 4 | 45, 000 | 578.2 |
| PINN (L-BFGS) | - | 1.69e - 3 | 1.69e - 4 | 22, 000 | 806.4 |
| FEM | 200, 000 | 5.26e - 9 | 1.37e - 9 | - | 4.7 |
|  | 400, 000 | 1.31e - 9 | 3.43e - 10 | - | 8.8 |
|  | 800, 000 | 3.29e - 10 | 8.57e - 11 | - | 18.1 |

Table 3. Comparison of error and time for solving one-dimensional viscous Burgers' equation[51].

| method | elements | $\Delta t$ | maximum error | rms error | epochs | training time/ wall-time(seconds) |
|---|---|---|---|---|---|---|
| DGM (Adam) | - | - | 4.57e - 2 | 5.76e - 3 | 128, 000 | 1797.8 |
| DGM (L-BFGS) | - | - | 7.50e - 3 | 1.55e - 3 | 28, 000 | 1813.5 |
| FEM | 2000 | 0.001 | 2.64e - 5 | 5.15e - 6 | - | 12.5 |
|  | 2000 | 0.0005 | 3.07e - 5 | 5.76e - 6 | - | 25.4 |
|  | 5000 | 0.001 | 1.89e - 5 | 1.74e - 6 | - | 26.0 |
|  | 5000 | 0.0005 | 4.13e - 6 | 7.90e - 7 | - | 50.8 |
|  | 10000 | 0.001 | 2.22e - 5 | 1.99e - 6 | - | 47.7 |
|  | 10000 | 0.0005 | 4.74e - 6 | 4.36e - 7 | - | 92.6 |

**(2) Poor generality**

Poor generality is also a major limitation of PINNs. On the one hand, it is difficult to handle complex engineering problems. On the other hand, even when faced with simple model equations, PINNs may fail to solve for various reasons.

Some typical examples include: for flow problems in fluid mechanics, the PDEs loss function in PINNs cannot guarantee a decrease in the total PDE residual of the entire computational domain under nonuniform collocation points, resulting in non-physical results[52] (as shown in Figure 3). For unsteady cylinder flow, due to the lack of disturbance in the solution process, PINNs are unable to obtain periodic vortex shedding (Karman vortex street)[53]. When the Reynolds number in the Navier Stokes equation is high, PINNs cannot obtain satisfactory results[54]. For turbulence simulation, the training of PINNs with turbulence models becomes more complex, the prediction results for the backward-facing step flow show significant differences compared to the reference solution[55]. When the solution of PDEs has multi-scale or high-frequency features, PINNs that follow the frequency principle[56] have insufficient approximation ability for small scale/high-frequency, making it difficult to obtain accurate results[57-59]. Wang et al.[57] used PINNs to solve the one-dimensional Poisson equation $\Delta u(x) = f(x)$, and its analytical solution was designed to have significant frequency discrepancy $u(x) = \sin(2\pi x) + 0.1\sin(50\pi x)$. The comparison between the NN predicted solution and the analytical solution is shown in Figure 4. We observe that the results of PINNs have significant errors. For time-dependent PDEs, PINNs that minimize the equation residual over the entire time interval violate the time evolution laws of dynamic systems, leading to failure in solving problems that are sensitive to initial conditions[60]. For problems where the external excitation is a point source (represented mathematically by a Dirac function), PINNs fail to solve correctly due to the singularity of the Dirac function[61].

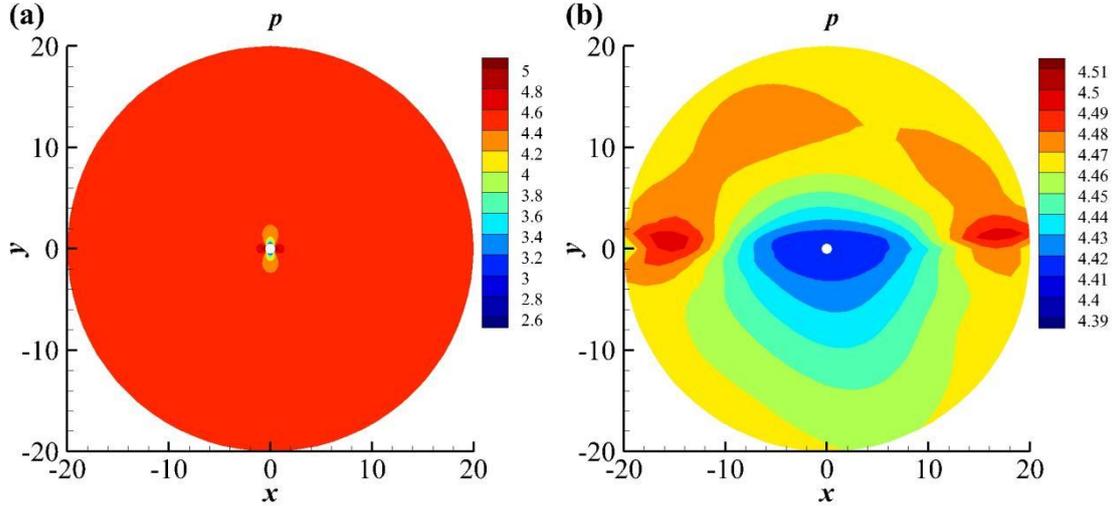

Figure 3. The comparison of pressure field between (a) the reference solution from finite volume method and (b) the solution given by PINNs for solving inviscid compressible flow over a circular cylinder[52].

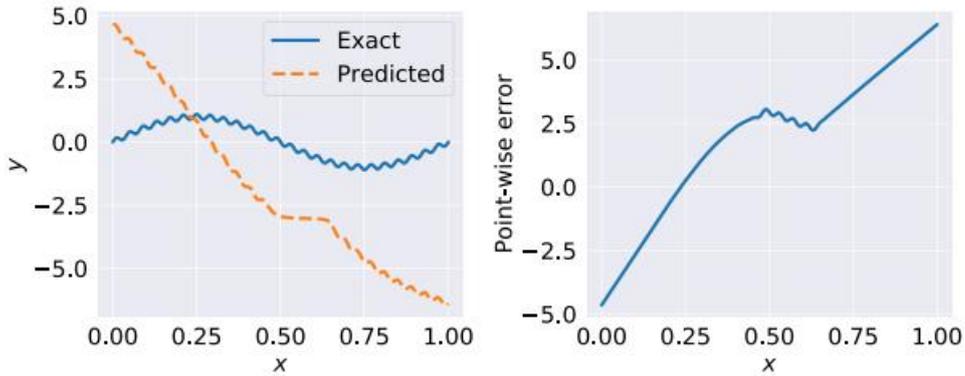

Figure 4. 1D Poisson equation: results obtained by training a physics-informed neural network. (a) Comparison of predicted and exact solutions. (b) Point-wise error between the predicted and the exact solution[57].

**(3) Numerous empirical parameters and strong parameter sensitivity**

The conflict of multiple losses has always been widely present in machine learning. To address this problem, researchers have introduced empirical parameters as loss weights to ensure a reasonable training process[62]. In PINNs, this issue is particularly prominent because there are not only conflicts between the IBCs, equation residuals, and observed data mismatch (inverse problem), but also conflicts between different equations (for equation systems) and equation residuals at different collocation points, all of which can undermine the correct solution of PINNs. For example, the conflict between PDE loss and boundary condition loss resulted in a significant deviation in the solution of the Helmholtz equation at the boundary[63], as shown in Figure 5. Conflicts in equation residuals at different times lead to a solution process that violates the time evolution laws of dynamical systems. Conflicts in equation residuals at collocation points with different sampling densities result in ill-conditioning[52]. Additionally, there are magnitude differences between the equation residuals involving differential terms and the losses from IBCs/observed data, which further exacerbate the difficulty of balancing the losses[58, 63-65].

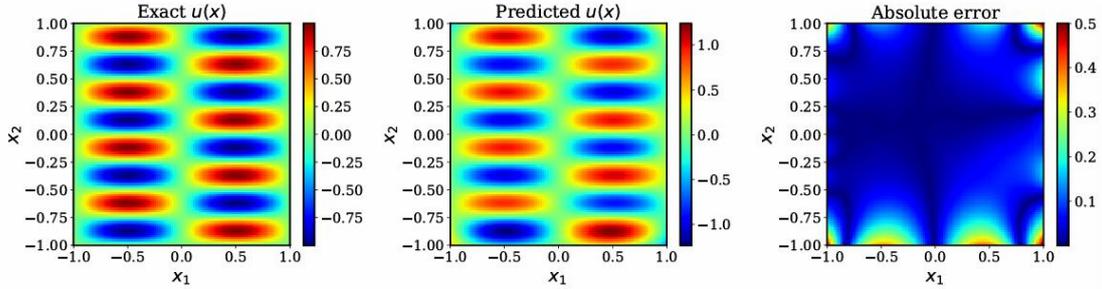

Figure 5. The solution of Helmholtz equation. (a) Exact solution. (b) The solution of PINNs. (c) Point-wise error[63].

To alleviate the loss conflict problem of PINNs, many researchers have conducted targeted research, and the common strategy is to weight the losses and equation residuals at each collocation point through empirical parameters. The weight design for various losses is mainly carried out through adaptive methods: Wang et al.[57] have designed a weight calculation method by statistically analyzing the gradient of the loss during the network training process. Yu et al.[66] have alleviated potential conflicts by projecting gradient descent directions for different losses. Xiang et al.[67] have constructed loss weights by incorporating maximum likelihood estimation. Wang et al.[58] have also designed a weight adaptive method by conducting neural tangent kernel (NTK) analysis on PINNs. Wang et al.[68] have proposed a loss balancing framework based on the concept of power-law scaling. For the weighting of equation residuals at each collocation point, Song et al.[52] have proposed volume weighting physics-informed neural networks (VW-PINNs), which weights equation residuals through the volume occupied by collocation points in the computational domain, effectively improving the performance of PINNs under nonuniform collocation points. In addition, some adaptive point weighting methods[69-71] have also been proposed to enhance vanilla PINNs frameworks. Undoubtedly, these methods have to some extent alleviated the conflict issues during the network training process. However, they have also introduced numerous empirical parameters with high algorithmic sensitivity, increasing the complexity of PINNs and potentially compromising robustness in solving process. In other words, these methods have not fundamentally eliminated conflicts of various losses.

Recently, researchers have proposed some methods of enforcing IBCs through hard constraints to eliminate the losses of IBCs in PINNs. These frameworks are mainly categorized into implementations involving fully-connected networks[44, 72-75] and convolutional networks[76-78]. For the former, the researchers constructed solution functions that a priori satisfy the IBCs by introducing boundary functions and distance functions. For the latter, the imposition of hard constraints is achieved through padding the graph, which is similar to the handling of classical numerical methods. These methods avoid the conflict between the loss of IBCs and other losses, but they still have limitations. On the one hand, conflicts still exist between different equations at different collocation points. On the other hand, the boundary function and distance function introduced in fully-connected networks are still empirical. In convolutional networks, PINNs cannot use automatic differentiation and rely on mesh, which results in PINNs losing their unique advantages over classical numerical methods, including solving high-dimensional problems[45, 79] and parametric problems[46, 80].

## 2 Exploration of the root causes of PINNs' failure in solving PDEs

The three main limitations of PINNs summarized above are based on the case results from relevant literature. In fact, this approach of summarizing is more from the perspective of PINNs users' operations. However, to improve PINNs, it is necessary to analyze the underlying mechanisms behind these limitations. Building on the three main limitations mentioned above, this section discusses the root causes of PINNs' failure in solving PDEs, analyzing the following three points: (1) Poor multiscale approximation ability and ill-conditioning caused by PDE losses; (2) Insufficient exploration of convergence and error analysis, resulting in weak mathematical rigor; (3) Inadequate integration of physical information, causing mismatch between residuals and iteration errors.

**(1) Poor multiscale approximation ability and ill-conditioning caused by PDE losses**

As is well known, searching for neural network parameters is an optimization problem aimed at minimizing the loss functions. According to the current research, whether it is data-driven modeling or solving PDEs, researchers commonly use deep neural networks with a large number of parameters to achieve sufficient nonlinear approximation capability. The optimization problem becomes high-dimensional and non-convex, making it extremely difficult to search for the global minimum. Compared to data-driven modeling, the optimization problem constructed by PINNs is more complex and challenging[64, 65].

The optimization difficulties of PINNs mainly come from two aspects. On the one hand, there are significant scale discrepancies such as multiple frequencies and strong anisotropy in the solution, which makes it difficult to be accurately approximated by the global basis functions constructed by vanilla PINNs. On the other hand, PDE losses lead to ill-conditioning in optimization. Compared to the $L_2$ regularization constraints in the data-driven modeling, the regularization constraints corresponding to PDE residuals composed of differential terms is non trivial. To handle the former difficulty, Liu et al.[81] and Wang et al.[57] both pointed out the challenges that PINNs face in the multiscale problems, and they designed improved frameworks by constructing multiscale inputs a priori and by scaling spatiotemporal coordinates in the sine space, respectively. In fact, these studies can be traced back to the work of Rahaman et al[82] and Xu et al.[56], who explained the fitting preferences of neural networks for low-frequency features from the perspectives of spectral bias and frequency domain. Cao et al.[83] have pointed out that the large gradient characteristics of inviscid airfoil flow near the wall in physical space are unfavorable for flow simulation, and have achieved satisfactory results in computational space with very few meshes through mesh transformation method. In contrast, the second-order finite volume method would require nearly 20 times more meshes to achieve the same accuracy, as shown in Figure 6. Wu et al.[84] have designed transformation parameters based on prior knowledge to perform linear transformations on PDEs, enhancing the learning ability of PINNs in the boundary layer region of the flow field. The author verified the effectiveness of the proposed method through a large number of examples.

For the latter difficulty, recent preprocessing studies on PINNs[85-87] have theoretically analyzed in detail the ill-conditioning in optimization caused by PDE loss. Rathore et al.[85] analyzed the condition numbers of the Hessian matrix corresponding to each loss term in PINNs (Figure 7), and pointed out the ill-conditioning in optimization caused by equation residuals from the perspective of the loss landscape. They further explained the advantages and limitations of the L-BFGS optimization algorithm in PINNs, and finally proposed the NysNewton CG (NNCG) optimization method to improve it. By introducing appropriate assumptions, De Ryck et al.[86] proved that the convergence rate of gradient descent in PINNs is related to the condition number of the composite operator composed of the Hermitian square of the differential operator and the kernel integral operator related to the tangent kernel of the underlying model, and proposed improvement strategies through operator preprocessing. Liu et al.[87] combined the ideas of traditional paradigms, transforming PDEs into a system of linear equations and analyzing the reasons for poor convergence and accuracy of PINNs from the perspective of coefficient matrix condition numbers, and proposed a method to preprocess equation residuals to improve algorithm performance. Cao et al.[88] established a relationship between the ill-conditioning of PINNs and the Jacobian matrix of the PDE system, demonstrating that the degree of ill-conditioning of PINNs is proportional to the condition number of the Jacobian matrix. They proposed an improved strategy based on the time-stepping method, successfully solving the forward problem of three-dimensional flow around a wing. Overall, despite PINNs embedding PDEs to ensure that network outputs better adhere to physics, satisfactory results can be achieved even with no data or only a small amount of data. However, the PDE loss also leads to poor trainability and even falling into ill-conditioning in optimization, which is a fundamental reason for the low accuracy, efficiency, and poor generality of PINNs.

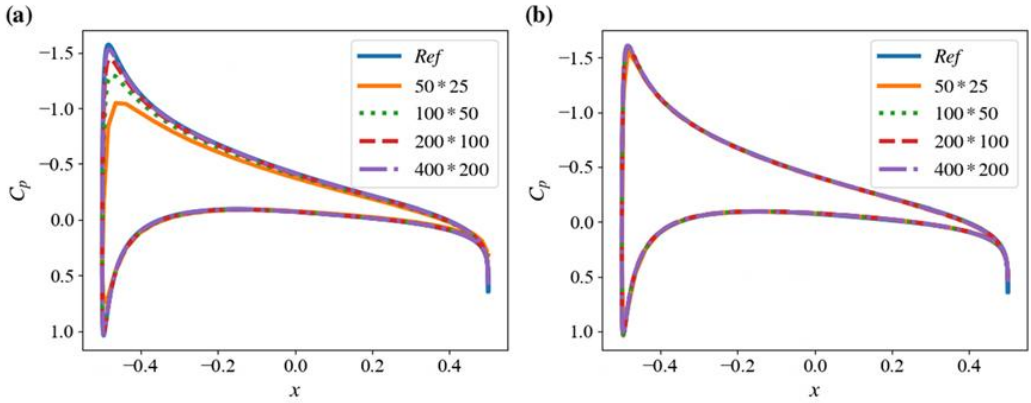

Figure 6. The pressure coefficient distributions of the flow around the NACA0012 airfoil obtained by (a) FVM and (b) NNfoil under different meshes[83].

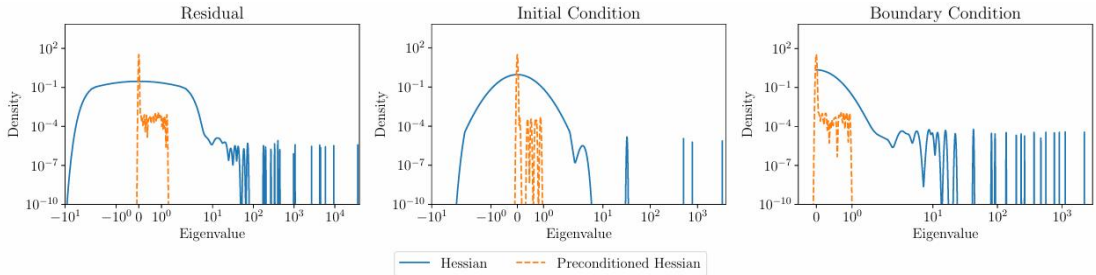

Figure 7. Spectral density of the Hessian and the preconditioned Hessian of each loss term for convection equation[85].

**(2) Insufficient exploration of convergence and error analysis, resulting in weak mathematical rigor**

Convergence is an important property considered in error analysis for classical numerical methods, reflecting whether the solution of the discretized equation approaches the solution of the differential equation as the spatial and temporal grid steps tend toward zero. In classical numerical methods, extensive research has focused on theoretical proofs of convergence and strategies for improving convergence[89-95], thereby enhancing the robustness of these methods in solving complex problems and engineering applications.

However, for PINNs, research on convergence remains to be deepened. Referring to the concept of classical numerical methods, the primary challenge in studying the convergence of PINNs is how to define the error of PINNs and the degrees of freedom that affect error magnitude. Regarding PINNs' error, Shin et al.[96] decompose the total error of PINNs into approximation error, estimation error, and optimization error from the perspective of statistical learning error. The approximation error refers to the theoretical error between the neural network of a given architecture and the true model, the estimation error arises when evaluating the model at unknown points, and the optimization error is the difference between the network's achieved accuracy and its theoretical accuracy after training. The approximation error depends solely on the complexity of the model architecture, while the estimation error and approximation error together form the generalization error. Optimization error originates from the current difficulty of finding the global minimum of a non-convex function. As for the degrees of freedom that affect error magnitude, they need to be defined according to the type of PINNs' error. The approximation error is related only to model architecture complexity, determined by the number of neurons. The universal approximation theorem indicates that a neural network with at least one hidden layer can approximate any function to arbitrary precision[97]. The number of neurons sets the precision the network can achieve, making it a degree of freedom that affects the approximation error. For estimation and optimization errors, the optimization process is influenced by multiple factors, such as model width and depth, dataset size, optimizers, training epochs, learning rate, and activation functions. A relatively effective research approach is to study the convergence characteristics of the error with changes in model width, model depth, and dataset size while keeping optimizers, training epochs, learning rate, and activation function constant. Then, by varying the optimizers, training epochs, learning rate, and activation functions, the consistency of the convergence behavior can be investigated[98].

Building on the error definitions for PINNs by Shin et al.[96], researchers have conducted theoretical studies on the convergence of PINNs[96, 99-102]. However, as noted in [98], current theoretical research on PINNs' convergence mainly focuses on analyzing generalization error bounds for different types of PDEs. By establishing a link between training error and generalization error, it aims to ensure that, under certain conditions, sufficiently low training error can imply a sufficiently low generalization error. Although this research has advanced the theoretical understanding of PINNs' convergence, these theories are based on the assumption that PINNs' optimization can reach a global minimum, which exceeds the capabilities of current optimizers. The neural tangent kernel (NTK) theory[58] considers the impact of the optimizers, exploring gradient descent convergence for fully connected neural networks in the infinite-width limit. By monitoring PINNs' convergence behavior through NTK and the relative change of the parameter vectors, results show that, for shallow networks in the infinite-width limit, gradient descent training with an infinitesimally small learning rate can guarantee convergence of PINNs training. However, the NTK theory's assumptions are idealized, requiring an infinitely wide single-layer network, low learning rates, and a gradient descent optimizer, which limits its practical applicability. Since current theoretical studies on PINNs' convergence are restricted to idealized cases and lack practical guidance, Pantidis et al.[98] investigated PINNs' convergence concerning network architecture complexity and dataset size through numerical studies. Considering randomness from initialization, Pantidis et al.[98] found that both training error and global error of PINNs tend to minimize with increasing network complexity and dataset size, as shown in Figure 8. Consistent convergence patterns were observed across different optimizers, training epochs, and learning rates. In summary, further progress in studying PINNs' convergence is needed through both theoretical and numerical approaches. On the one hand, theoretical research should enhance generality and applicability to provide practical guidance; on the other hand, numerical research should expand, verifying and generalizing findings across different types of problems.

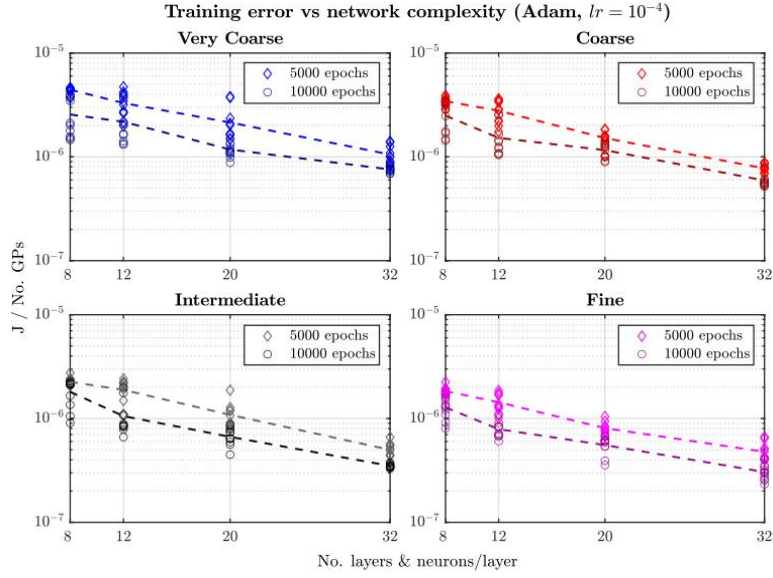

(a) Variation of training error with respect to network complexity.

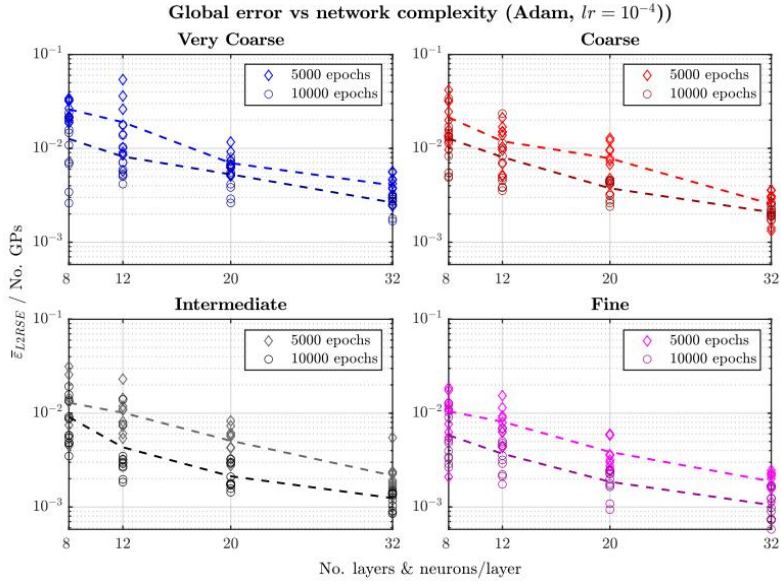

(b) Variation of global error with respect to network complexity.

Figure 8. The (a) training error and (b) global error of PINNs tend to minimize as network complexity increases (using the Adam optimizer with a learning rate of $10^{-4}$, with results recorded on four different grid sets)[98].

**(3) Inadequate integration of physical information, causing mismatch between residuals and iteration errors**

Residuals and iterative errors are two important concepts in the CFD simulation.

Residuals refer to the degree to which the governing equation is satisfied at each discrete point during a numerical iteration process. In other words, residuals measure the imbalance between the current solution and the governing equation. While iterative errors refer to the difference between the current numerical solution and the exact solution (theoretical solution of the discrete equations). Iterative errors reflect the process of the computed solution gradually approaching the theoretical solution. The relationship between the two is that small iterative errors usually require small residuals as a basis, but conversely, small residuals do not always lead to small iterative errors[103].

In PINNs, the mismatch between residuals and iterative errors is further exacerbated due to the inherent characteristics of the method. The vanilla PINNs have two typical characteristics: one is the use of soft constraints on boundary conditions, and the other is the discretization of the solution space using scattered points. These features give PINNs flexibility in handling various boundary condition types and geometric shapes. From the perspective of the loss function, the solving process in PINNs involves progressively minimizing the equation loss on scattered points and the boundary condition loss on boundary points, with the solution deemed complete once the residuals are below a certain threshold. However, from a physical perspective, the soft constraints on boundaries can lead to deviations in boundary conditions, potentially affecting the well-posedness of the PDE solutions. Additionally, an improper distribution of scattered points may result in mismatches between the characteristics of the PDE in continuous and discrete spaces.

Let's start by analyzing the treatment of boundary conditions. Classical numerical methods handle boundary conditions by explicitly applying values, gradients, or a combination of both at the boundarie[104-106]. These imposed boundary conditions are incorporated into the algebraic equation system formed by the numerical discretization of the PDEs, directly influencing values within the solution domain through iterative solving. In classical methods, boundary handling is a strict physical constraint; though there may be variations in accuracy across different boundary handling methods, the physical characteristics of the problem are directly transferred across the entire solution domain via the boundary. However, in PINNs, boundary conditions are addressed as part of the loss function, which approximates boundary conditions by minimizing the difference between model predictions and actual boundary conditions. This means boundary conditions are applied in a statistical rather than a strict physical sense. Soft boundary constraints make it challenging for complex boundary conditions to be strictly satisfied in PINNs. Although the numerical boundary error magnitude may be small, such an error can lead to significant deviations in boundary conditions from a physical perspective. For instance, unlike flow around bluff bodies, flow around airfoils may be sensitive to boundary condition errors, which implicitly alter the airfoil shape distribution; even slight deviations in airfoil shape can have a substantial impact on flow behavior[107, 108]. Designing appropriate loss function weight coefficients based on the physical characteristics of the problem to ensure boundary condition errors meet accuracy requirements is key to improving the consistency of PINNs' results with expected physical field behavior.

Next, let's analyze from the perspective of domain discretization. Classical numerical methods generally discretize the solution domain using structured or unstructured grids. The grid distribution is often pre-adjusted according to the physical characteristics of the problem, such as locally refining around boundary layers or shock regions in fluid flow. This targeted handling of grids not only reduces computation cost but also incorporates the physical scale features of the problem into the grid distribution, enhancing the capability of classical numerical methods to handle complex problems[109, 110]. In contrast, PINNs represent the solution domain using scattered points for discretization. Although scattered points can better handle complex geometries without needing to consider grid topology, a scattered point distribution that does not align with the physical spatial scale features increases the difficulty for vanilla PINNs to learn the physical characteristics of the solution domain. Ideally, the distribution of points in PINNs should be associated with the physical features of the domain. Adaptive approaches, for example, could be used to adjust the point distribution in PINNs[71, 111-113]. Moreover, grids have the advantage of containing volume information, a physical attribute that, when introduced into the loss function of PINNs, can positively influence the approximation of the optimization objective[52], as shown in Figure 9.

In summary, PINNs require the integration of domain-specific expert knowledge to effectively design and optimize the model, enhancing the physical relevance of the model and increasing its applicability in complex scenarios.

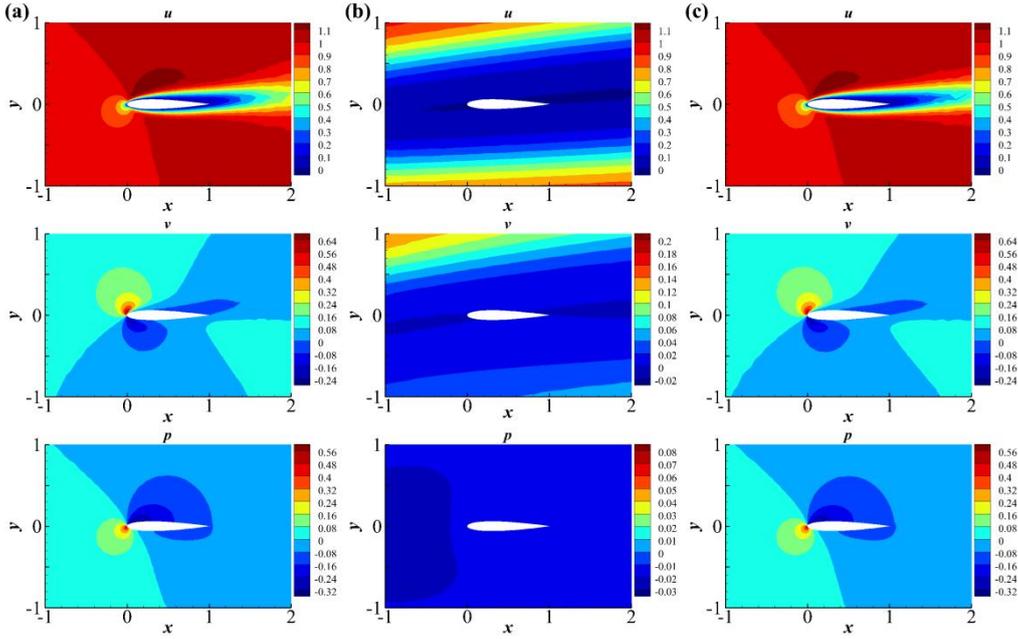

Figure 9. Comparison of the near-wall flow field around the viscous incompressible NACA0012 airfoil: (a) reference solution; (b) PINNs; (c) VW-PINNs[52].

**3 Summary and Outlook on Solving PDEs with Neural Networks**

PINNs introduce PDEs into the loss function of neural networks, offering a novel approach to solving PDEs through optimization, which has driven their widespread application across various fields and suggests a promising future for this research direction. However, achieving stable training and obtaining correct results remain a challenge in many complex scenarios. This paper analyzes the case results of PINNs from relevant literature, summarizing the failure phenomena and limitations of PINNs: (1) Low solution accuracy and high computation cost; (2) Poor generality; (3) Numerous empirical parameters and strong parameter sensitivity. An in-depth analysis of these limitations is then conducted, revealing the underlying mechanisms that contribute to these constraints. Three root causes for the failure of PINNs in solving complex problems are identified: (1) Poor multiscale approximation ability and ill-conditioning caused by PDE losses; (2) Insufficient exploration of convergence and error analysis, resulting in weak mathematical rigor; (3) Inadequate integration of physical information, causing mismatch between residuals and iteration errors. The above analysis provides a basis for further improvements in PINNs. We believe that subsequent improvements in using neural networks to solve PDEs can focus on the following aspects.

**(1) Analysis of ill-conditioning in PINNs and mitigation strategies**

The ill-conditioning during optimization in PINNs is a major obstacle to their application in complex engineering problems. Different perspectives on the ill-conditioning of PINNs have inspired various mitigation strategies. One perspective is to relate the ill-conditioning of PINNs to the Hessian matrix of the loss function. Rathore et al.[85] explained the optimization stagnation caused by residual loss in terms of the spectral density of the Hessian matrix of the loss function and improved PINNs performance by enhancing the gradient descent algorithm. De Ryck et al.[86] argued that the convergence rate of gradient descent in PINNs is related to the condition number of a composite operator, which consists of the Hermitian square of differential operators and the kernel integral operator associated with the tangent kernel for the underlying model. They proposed an improvement strategy from the perspective of operator preconditioning. Another perspective is to relate the ill-conditioning of PINNs to the Jacobian matrix of the PDE system. Based on an understanding of ill-conditioning in classical numerical methods, Cao et al.[88] associated the ill-conditioning of PINNs with the Jacobian matrix of the PDE system rather than the Hessian matrix of the loss function. They constructed a controlled system based on control theory for any dynamic system, allowing adjustment of the gain to alter the Jacobian matrix's condition number. Numerical examples then verified the correlation between the ill-conditioning of PINNs and that of the PDE system's Jacobian matrix. This evolving understanding not only sheds light on the intrinsic limitations of PINNs but also opens avenues for their broader application in solving complex physical problems.

**(2) Improvements to PINNs by enforcing temporal causality**

When solving time-dependent PDE problems, PINNs typically select residual sampling points across the entire spacetime domain and use all these points simultaneously for training the neural network parameters. This solving process indicates that PINNs treat the spacetime dimensions of PDEs in a uniform manner. Clearly, this disrupts the inherent temporal causality of the PDE system and weakens the physical validity of the PINNs solution. Enforcing temporal causality is an improvement direction for PINNs, ensuring the correct representation of the intrinsic physical properties of unsteady systems as they evolve over time.

To adhere to the physical evolution of flow over time, Wang et al.[60] assigned weights to the collocation points along the time dimension, only solving for the next time step when the loss at the previous time step is sufficiently low, indicating that the solution has been adequately obtained. Krishnapriyan et al.[64] performed time segmentation with equal intervals and solved each time interval sequentially. Cao et al.[54, 88] proposed the time-stepping-oriented neural network (TSONN), which replaces the original equation with a series of pseudo-time stepping equations. This approach aligns the convergence history of the neural network with the pseudo-time evolution of the physical system, making it applicable to both time-dependent and time-independent problems. These studies demonstrate that adhering to temporal causality can significantly improve the performance of PINNs in solving problems.

**(3) Empowering PINNs with classical numerical methods**

Classical numerical methods have made significant progress over several decades, accumulating a profound understanding of physics in areas such as mesh generation strategies and numerical scheme design. How to leverage various techniques from classical numerical methods and their deep insights into physics to empower the solution of PINNs, while not losing the core advantages that PINNs offer, is a direction worth considering for improving PINNs. He et al.[114], inspired by the entropy viscosity method[115] developed in computational fluid dynamics (CFD), proposed the artificial viscosity (AV)-based PINN method to stabilize the simulation of high Reynolds number flows, successfully solving the two-dimensional lid-driven cavity flow at Re = 1000. Similarly, the TSONN method proposed by Cao et al.[54, 88] was inspired by the pseudo-time stepping method in CFD and achieved simulated lid-driven cavity flow and three-dimensional flow around a wing at Re = 5000. In classical numerical methods, the grid distribution also plays a crucial role. For instance, in high Reynolds number wall-bounded turbulence with multiscale phenomenon, anisotropic grids within the boundary layer are often indispensable. Drawing from grid transformation strategies in CFD, Cao et al.[83] and Zhang et al.[116] used neural networks to learn flows in the computational space rather than the physical space. By partially incorporating the physical scale characteristics into the grid distribution, they significantly alleviated challenges in using PINNs to solve inviscid airfoil flow and high Reynolds number wall-bounded turbulence problems. In the future, the integration of more numerical methods with PINNs holds significant potential for exploration and development. This combination could leverage the strengths of traditional numerical techniques and neural networks, addressing limitations in current methods while enhancing the efficiency and accuracy of solving complex physical problems.